\documentstyle[prd,aps,preprint,tighten,epsfig]{revtex}

\begin{document}

\draft

\title{Radiative Corrections to Neutrino Mixing and CP Violation \\
in the Minimal Seesaw Model with Leptogenesis}
\author{{\bf Jian-wei Mei} ~ and ~ {\bf Zhi-zhong Xing}}
\address{CCAST (World Laboratory), P.O. Box 8730, Beijing 100080, China \\
and Institute of High Energy Physics, Chinese Academy of Sciences, \\
P.O. Box 918 (4), Beijing 100039, China
\footnote{Mailing address} \\
({\it Electronic address: jwmei@mail.ihep.ac.cn; xingzz@mail.ihep.ac.cn}) }
\maketitle

\begin{abstract}
Radiative corrections to neutrino mixing and CP violation are analyzed in
the minimal seesaw model with two heavy right-handed neutrinos. We find
that textures of the effective Majorana neutrino mass matrix
are essentially stable against renormalization effects. Taking account
of the Frampton-Glashow-Yanagida ansatz for the Dirac neutrino Yukawa
coupling matrix, we calculate the running effects of light neutrino
masses, lepton flavor mixing angles and CP-violating phases for both
$m_1 =0$ (normal mass hierarchy) and $m_3 =0$ (inverted mass hierarchy)
cases in the standard model and in its minimal supersymmetric extension.
Very instructive predictions for the cosmological baryon number asymmetry
via thermal leptogenesis are also given with the help of low-energy
neutrino mixing quantities.
\end{abstract}


\newpage

\section{Introduction}

In the standard model (SM), lepton number conservation is assumed
and neutrinos are exactly massless Weyl particles. However, the recent
Super-Kamiokande \cite{SK}, SNO \cite{SNO}, KamLAND \cite{KM} and
K2K \cite{K2K} experiments have provided us with very compelling evidence
that neutrinos are actually massive and lepton flavors are mixed. The most
economical modification of the SM, which can
both accommodate neutrino masses and allow lepton number violation to
explain the cosmological baryon asymmetry via leptogenesis \cite{FY},
is to introduce two heavy right-handed neutrinos $N_{1,2}$ and
keep the Lagrangian of electroweak interactions invariant under
$\rm SU(2)_L \times U(1)_Y$ gauge
transformation \cite{FGY,Tanimoto,GX03,MSM}
In this case, the Yukawa interactions of leptons are described by
\begin{equation}
-{\cal L}_{\rm Y(SM)} \; =\; \bar{l}_{\rm L} Y_l e^{~}_{\rm R} H
+ \bar{l}_{\rm L} Y_\nu \nu^{~}_{\rm R} H^{\rm c} +
\frac{1}{2} \overline{\nu^{\rm c}_{\rm R}} M_{\rm R} \nu^{~}_{\rm R}
+ {\rm h.c.} \; ,
%       (1)
\end{equation}
where $l_{\rm L}$ denotes the left-handed lepton doublet; $e^{~}_{\rm R}$
and $\nu^{~}_{\rm R}$ stand respectively for the right-handed charged
lepton and Majorana neutrino singlets; and $H$ is the Higgs-boson
weak isodoublet (with $H^{\rm c} \equiv i\sigma^{~}_2 H^*$). If the
minimal supersymmetric standard model (MSSM) is taken into account,
one may similarly write out the Yukawa interactions of leptons:
\begin{equation}
-{\cal L}_{\rm Y(MSSM)} \; =\; \bar{l}_{\rm L} Y_l e^{~}_{\rm R} H_1
+ \bar{l}_{\rm L} Y_\nu \nu^{~}_{\rm R} H_2 +
\frac{1}{2} \overline{\nu^{\rm c}_{\rm R}} M_{\rm R} \nu^{~}_{\rm R}
+ {\rm h.c.} \; ,
%       (2)
\end{equation}
where $H_{1,2}$ (with hypercharges $\pm 1/2$) are the MSSM Higgs doublets.

Without loss of generality, both the heavy Majorana
neutrino mass matrix $M_{\rm R}$ and the charged-lepton Yukawa
coupling matrix $Y_l$ can be taken to be diagonal, real and positive.
In this specific flavor basis, the Dirac neutrino Yukawa coupling matrix
$Y_\nu$ is a complex $3\times 2$ rectangular matrix:
\begin{equation}
Y_\nu \; =\; \left ( \matrix{
a_1 & a_2 \cr
b_1 & b_2 \cr
c_1 & c_2 \cr} \right ) \; .
%       (3)
\end{equation}
Below the mass scale of the lightest right-handed neutrino $N_1$
(denoted as $M_1$), $M_{\rm R}$ can be integrated out of the theory.
Such a treatment corresponds to a replacement of
the last two terms in ${\cal L}_{\rm Y}$ by a dimension-5 operator,
whose coupling matrix takes the seesaw \cite{SS} form
\begin{eqnarray}
\kappa^{~} (M_1) \; = \; Y_\nu M^{-1}_{\rm R} Y^T_\nu
\; =\; \left ( \matrix{
\displaystyle\frac{a^2_1}{M_1} &
\displaystyle\frac{a_1 b_1}{M_1} &
\displaystyle\frac{a_1 c_1}{M_1} \cr\cr
\displaystyle\frac{a_1 b_1}{M_1} &
\displaystyle\frac{b^2_1}{M_1} &
\displaystyle\frac{b_1 c_1}{M_1} \cr\cr
\displaystyle\frac{a_1 c_1}{M_1} &
\displaystyle\frac{b_1 c_1}{M_1} &
\displaystyle\frac{c^2_1}{M_1} \cr}
\right ) +
\left ( \matrix{
\displaystyle\frac{a^2_2}{M_2} &
\displaystyle\frac{a_2 b_2}{M_2} &
\displaystyle\frac{a_2 c_2}{M_2} \cr\cr
\displaystyle\frac{a_2 b_2}{M_2} &
\displaystyle\frac{b^2_2}{M_2} &
\displaystyle\frac{b_2 c_2}{M_2} \cr\cr
\displaystyle\frac{a_2 c_2}{M_2} &
\displaystyle\frac{b_2 c_2}{M_2} &
\displaystyle\frac{c^2_2}{M_2} \cr}
\right ) \; .
%       (4)
\end{eqnarray}
It has been shown that ${\rm Det}[\kappa^{~} (M_1)] =0$
holds for arbitrary $a_i$, $b_i$ and $c_i$ \cite{X03}.
This is a very special feature of the minimal seesaw model.

After spontaneous gauge symmetry breaking, the neutral component of
$H$ acquires the vacuum expectation value $v \approx 174$ GeV. Then
one obtains the charged lepton mass matrix $M_l = v Y_l (M_Z)$ and
the light (left-handed) Majorana neutrino mass matrix
$M_\nu = v^2 \kappa (M_Z)$ at the electroweak scale $\mu = M_Z$
in the SM. The neutral
components of $H_1$ and $H_2$ may similarly acquire the vacuum
expectation values $v\cos\beta$ and $v\sin\beta$ at the electroweak
symmetry breaking scale. It turns out that
$M_l = v\cos\beta Y_l (M_Z)$ and
$M_\nu = v^2 \sin^2\beta \kappa (M_Z)$ in the MSSM. Note that
$\kappa (M_Z)$ and $\kappa (M_1)$ can be related to each other via
the following one-loop renormalization-group equation (RGE) \cite{RGE}:
\begin{equation}
16\pi^2 \frac{{\rm d} \kappa}{{\rm d} t} \; = \;
\alpha \kappa + C \left [ \left (Y_l Y^\dagger_l \right ) \kappa
+ \kappa \left (Y_l Y^\dagger_l \right )^T \right ] \; ,
%       (5)
\end{equation}
where $t = \ln (\mu/M_1)$ with $\mu$ being the renormalization
scale. In the SM \cite{RGE1} or in its minimal supersymmetric
extension \cite{RGE2},
\begin{eqnarray}
C_{\rm SM} & = & -\frac{3}{2} \; ,
\nonumber \\
\alpha^{~}_{\rm SM} & = & -3g^2_2 + 6 f^2_t + \lambda \; ~~~~~~~
%       (6)
\end{eqnarray}
or
\begin{eqnarray}
C_{\rm MSSM} & = & 1 \; ,
\nonumber \\
\alpha^{~}_{\rm MSSM} & = & -\frac{6}{5} g^2_1 - 6g^2_2 + 6 f^2_t \; ,
%       (7)
\end{eqnarray}
where $g^{~}_{1,2}$ denote the gauge couplings, $f_t$ denotes
the top-quark Yukawa coupling, and $\lambda$ denotes the
Higgs self-coupling in the SM
%%%%%%%%%%%%%%%%%%%%%%
\footnote{In the expression of $\alpha^{~}_{\rm SM}$, we have neglected
very small contributions from the lighter quarks and charged leptons.
Similarly, the up- and charm-quark Yukawa couplings have been neglected
in the expression of $\alpha^{~}_{\rm MSSM}$.}.
%%%%%%%%%%%%%%%%%%%%%%
If $M_1 \gg M_Z$ holds, some deviation of $\kappa (M_Z)$ from
$\kappa (M_1)$ must take place.

Apparently, it is $\kappa (M_Z)$ or $M_\nu$
that governs the low-energy phenomenology of neutrino masses and
lepton flavor mixing, because $Y_l$ keeps diagonal in the RGE running
from $M_1$ to $M_Z$. In all recent analyses of the minimal seesaw
model \cite{FGY,Tanimoto,GX03,MSM}, however, the quantum corrections
to $\kappa$ at the electroweak scale have been neglected for the sake
of simplicity and illustration. The importance of RGE effects on the
evaluation of baryogenesis via leptogenesis in the {\it bottom-up}
approach (i.e., from low energies to the mass scale of heavy
right-handed neutrinos) has been pointed out by some authors \cite{L},
but a careful analysis of such effects in the minimal seesaw model
has not been done.

The purpose of this paper is to examine the stability of $\kappa$ against
radiative corrections in the minimal seesaw model.
We find that the texture of $\kappa$ is essentially
stable in the RGE evolution from $M_1$ to $M_Z$. To be specific, we
calculate the running effects of neutrino masses, lepton flavor mixing
angles and CP-violating phases by taking account of the
Frampton-Glashow-Yanagida (FGY) ansatz for $Y_\nu$ \cite{FGY}.
The cosmological baryon number asymmetry via leptogenesis is also
calculated at the scale $\mu = M_1$ with the help of
low-energy neutrino mixing quantities.

\section{RGE running effects from $M_1$ to $M_Z$}

In the flavor basis chosen above, one may simplify the RGE in Eq. (5)
and get the radiative corrections to $\kappa$ at $M_Z$. Let us define
the evolution functions
\begin{eqnarray}
I_\alpha & = & \exp \left [- \frac{1}{16\pi^2} \int^{\ln (M_1/M_Z)}_0
\alpha (t) ~ {\rm d} t \right ] \; ,
\nonumber \\
I^{~}_l & = & \exp \left [- \frac{C}{16\pi^2} \int^{\ln (M_1/M_Z)}_0
f^2_l (t) ~ {\rm d} t \right ] \; ,
%       (8)
\end{eqnarray}
where $f^{~}_l$ (for $l=e,\mu,\tau$) denote the Yukawa coupling
eigenvalues of charged leptons. Then we solve Eq. (5) and arrive at
\begin{equation}
\kappa (M_Z) \; =\;  I_\alpha
\left ( \matrix{
I_e & 0 & 0 \cr
0 & I_\mu & 0 \cr
0 & 0 & I_\tau \cr} \right )
\kappa (M_1)
\left ( \matrix{
I_e & 0 & 0 \cr
0 & I_\mu & 0 \cr
0 & 0 & I_\tau \cr} \right ) \; .
%       (9)
\end{equation}
The overall factor $I_\alpha$ only affects the magnitudes of light
neutrino masses, while $I^{~}_l$ (for $l=e,\mu,\tau$) can modify
both neutrino masses and lepton flavor mixing parameters
\cite{Ellis}. The strong mass hierarchy of three charged leptons
(i.e., $f_e < f_\mu < f_\tau$) implies that $I_e < I_\mu < I_\tau$
(SM) or $I_e > I_\mu > I_\tau$ (MSSM) holds below the scale $M_1$
.

Two comments on the consequences of Eq. (9) are in order.

(1) The determinant of $\kappa$, which vanishes at the scale $M_1$,
keeps vanishing at the scale $M_Z$. This point can clearly be seen from
the relation
\begin{equation}
{\rm Det}[\kappa (M_Z)] \; =\; I^3_\alpha I^2_e I^2_\mu I^2_\tau ~
{\rm Det}[\kappa (M_1)] \; .
%       (10)
\end{equation}
Because of $\left |{\rm Det}[\kappa (M_Z)] \right | = m_1 m_2 m_3$,
where $m_i$ (for $i=1,2,3$) denote the masses of three light neutrinos,
one may conclude that one of the three neutrino masses must vanish.
Considering that $m_2 > m_1$ is required by current solar neutrino
oscillation data \cite{SNO}, we are left with either $m_1=0$ (normal
mass hierarchy) or $m_3 =0$ (inverted mass hierarchy).

(2) Comparing between Eqs. (4) and (9), we find that the radiative
correction to $\kappa$ can effectively be expressed as the RGE running
effects in $a_i$, $b_i$ and $c_i$ of $Y_\nu$ (in the assumption that
$M_1$ keeps unchanged):
\begin{eqnarray}
a_i (M_Z) & = & I_e \sqrt{I_\alpha} ~ a_i (M_1) \; ,
\nonumber \\
b_i (M_Z) & = & I_\mu \sqrt{I_\alpha} ~ b_i (M_1) \; ,
\nonumber \\
c_i (M_Z) & = & I_\tau \sqrt{I_\alpha} ~ c_i (M_1) \; ,
%       (11)
\end{eqnarray}
where $i=1$ or $2$. These simple relations imply that possible texture
zeros of $\kappa$ at $M_1$ remain the same at $M_Z$,
at least at the one-loop level of RGE evolution. Hence the texture
of $\kappa$ is essentially stable against quantum corrections from
$M_1$ to $M_Z$.

For illustration, we typically take $m_t (M_Z) \approx 181$ GeV \cite{Koide}
to calculate the evolution functions $I_\alpha$ and $I^{~}_l$ (for
$l=e,\mu,\tau$). It is found that $I_e \approx I_\mu \approx 1$ is an
excellent approximation both in the SM and in the MSSM. Thus the RGE
running of $\kappa$ is mainly governed by $I_\alpha$ and $I_\tau$.
The behaviors of $I_\alpha$ and $I_\tau$ changing with $M_1$ are shown
in Fig. 1. One can see that $I_\tau \approx 1$ is also a good
approximation, in particular in the SM. Hence the evolution of light
neutrino masses must be dominated by $I_\alpha$, which may significantly
deviate from unity in the SM with reasonable values of the Higgs mass
$m^{~}_H$ and in the MSSM with large values of $\tan\beta$.

The effect of $I_\tau$ in the running of $\kappa$ will certainly lead to the
evolution of the lepton flavor mixing matrix $V$. At the electroweak scale,
$V$ is defined to diagonalize the neutrino mass matrix
$M_\nu$ in the chosen flavor basis; i.e.,
$V^\dagger M_\nu V^* = {\rm Diag} \{m_1, m_2, m_3 \}$. A commonly-used
parametrization of $V$ is
\begin{equation}
V \; = \; \left ( \matrix{
c_x c_z & s_x c_z & s_z \cr
- c_x s_y s_z - s_x c_y e^{-i\delta} &
- s_x s_y s_z + c_x c_y e^{-i\delta} &
s_y c_z \cr
- c_x c_y s_z + s_x s_y e^{-i\delta} &
- s_x c_y s_z - c_x s_y e^{-i\delta} &
c_y c_z \cr } \right )
\left ( \matrix{
e^{i\rho} & 0 & 0 \cr
0 & e^{i\sigma} & 0 \cr
0 & 0 & 1 \cr} \right ) \; ,
%       (12)
\end{equation}
where $s_x \equiv \sin\theta_x$, $c_x \equiv \cos\theta_x$, and so
on. The Dirac phase $\delta$ measures CP violation in normal
neutrino oscillations, while the Majorana phases $\rho$ and
$\sigma$ are relevant to the neutrinoless double beta decay
\cite{FX01}. Current data on solar, atmospheric and reactor
neutrino oscillations yield $\theta_x \approx 32^\circ$ and
$\theta_y \approx 45^\circ$ (best-fit values \cite{FIT}) as well
as $\theta_z < 12^\circ$ \cite{CHOOZ}. The typical mass-squared
differences of solar and atmospheric neutrino oscillations read as
$\Delta m^2_{\rm sun} \equiv m^2_2 - m^2_1 \approx 7.13 \times
10^{-5} ~{\rm eV}^2$ and $\Delta m^2_{\rm atm} \equiv |m^2_3 -
m^2_2| \approx 2.6 \times 10^{-3} ~{\rm eV}^2$ (best-fit values
\cite{FIT}), respectively. Following Ref. \cite{Casas} and Ref.
\cite{Lindner}, one may derive the RGEs of $(m_1, m_2, m_3)$,
$(\theta_x, \theta_y, \theta_z)$ and $(\delta,\rho, \sigma)$ with
the help of Eq. (5). The relevant analytical results can
considerably be simplified, if the smallness of $s_z$ and $\Delta
m^2_{\rm sun}/\Delta m^2_{\rm atm}$ is taken into account. It
should be noted that the parametrization of $V$ used here is
somehow different from that taken in Ref. \cite{Lindner}. To be
complete and explicit, we present our approximate analytical
expressions of $(\dot{m}_1, \dot{m}_2, \dot{m}_3)$,
$(\dot{\theta}_x, \dot{\theta}_y, \dot{\theta}_z)$ and
$(\dot{\delta},\dot{\rho}, \dot{\sigma})$ in Appendix A. Because
$m_1 =0$ or $m_3 =0$ must hold in the minimal seesaw model under
consideration, it is actually possible to obtain much simpler
results from Eqs. (A1)--(A4).

\section{FGY ansatz and radiative corrections}

The minimal seesaw model itself has no restriction on the structure
of $Y_\nu$. In Ref. \cite{FGY}, Frampton, Glashow and Yanagida (FGY)
have conjectured that $Y_\nu$ may have two texture zeros:
$a_2 = c_1 =0$ or $a_2 = b_1 =0$, which could stem from an underlying
horizontal flavor symmetry. The texture of $\kappa (M_1)$ in Eq. (4)
can then be simplified:
\begin{equation}
\kappa (M_1) \; =\; \left ( \matrix{
\displaystyle \frac{a^2_1}{M_1} &
\displaystyle \frac{a_1 b_1}{M_1} &
{\bf 0} \cr\cr
\displaystyle \frac{a_1 b_1}{M_1} &
\displaystyle \frac{b^2_1}{M_1} + \frac{b^2_2}{M_2} &
\displaystyle \frac{b_2 c_2}{M_2} \cr\cr
{\bf 0} &
\displaystyle \frac{b_2 c_2}{M_2} &
\displaystyle \frac{c^2_2}{M_2} \cr} \right ) \;
%       (13)
\end{equation}
in the $a_2 = c_1 =0$ case; or
\begin{equation}
\kappa (M_1) \; =\; \left ( \matrix{
\displaystyle \frac{a^2_1}{M_1} &
{\bf 0} &
\displaystyle \frac{a_1 c_1}{M_1} \cr\cr
{\bf 0} &
\displaystyle \frac{b^2_2}{M_2} &
\displaystyle \frac{b_2 c_2}{M_2} \cr\cr
\displaystyle \frac{a_1 c_1}{M_1} &
\displaystyle \frac{b_2 c_2}{M_2} &
\displaystyle \frac{c^2_1}{M_1} + \frac{c^2_2}{M_2} \cr} \right ) \;
%       (14)
\end{equation}
in the $a_2 = b_1 = 0$ case. Note that the $a_1 = c_2 = 0$ and
$a_1 = b_2 = 0$ cases are respectively equivalent to the
$a_2 = c_1 = 0$ and $a_2 = b_1 = 0$ cases in phenomenology. Note
also that the $b_2 = c_1 = 0$ (or $b_1 = c_2 = 0$) case, in which
the (2,3) and (3,2) matrix elements of $\kappa (M_1)$ vanish,
is not favored by current neutrino oscillation data \cite{X03} and
will not be taken into account in the subsequent discussions. In
addition, the two instructive patterns of $\kappa (M_1)$ in Eqs. (13) and
(14) are expected to have very similar consequences on neutrino masses,
lepton flavor mixing angles and CP-violating phases at low-energy scales
(See Ref. \cite{GX03} for some detailed discussions). We shall
therefore concentrate only on the FGY ansatz given in Eq. (13) later on.

It is worth remarking that $a_1$, $b_{1,2}$ and $c_2$ in Eq. (13)
are all complex parameters. Given the specific parametrization of $V$
in Eq. (12), it is straightforward to obtain
\begin{equation}
\kappa (M_Z) \; =\; \frac{M_\nu}{\Omega} \; =\; V \left ( \matrix{
\displaystyle\frac{m_1}{\Omega} & 0 & 0 \cr
0 & \displaystyle\frac{m_2}{\Omega} & 0 \cr
0 & 0 & \displaystyle\frac{m_3}{\Omega} \cr} \right ) V^T \; ,
%       (15)
\end{equation}
where $\Omega \equiv v^2$ in the SM and $\Omega \equiv v^2 \sin^2\beta$
in the MSSM. Then the phases of $a_1$, $b_{1,2}$ and $c_2$ can be
determined in terms of $\delta$, $\rho$ and $\sigma$.
In Refs. \cite{FGY,GX03}, the phase convention
$\arg (a_1) = \arg (b_2) = \arg (c_2) =0$ has
been taken. We do not adopt this phase convention in the present work.
The physical results predicted by the FGY ansatz are certainly
independent of any specific phase convention.

Note that $\kappa (M_Z)$ takes the same texture as $\kappa (M_1)$,
and their corresponding matrix elements are related to each other
via Eq. (11). This observation implies
that the {\it bottom-up} approach should be more convenient for the
numerical evaluation of radiative corrections --- namely, we determine the
parameters of the FGY ansatz at low energies by using current neutrino
oscillation data, and then run them to the mass scale $M_1$
to examine how large the renormalization effects are.

\subsection{Normal neutrino mass hierarchy ($m_1 =0$)}

It is rather obvious that $m_1 =0$ leads to
$m_2 =\sqrt{\Delta m^2_{\rm sun}} \approx 8.4 \times 10^{-3}$ eV and
$m_3 =\sqrt{\Delta m^2_{\rm sun} + \Delta m^2_{\rm atm}}
\approx 5.2 \times 10^{-2}$ eV. Furthermore, $m_1 =0$ implies that
only the Majorana phase $\sigma$ is physically meaningful. Taking
account of the texture zero $(M_\nu)_{13} =0$, one can determine
both $\delta$ and $\sigma$ in terms of the flavor mixing angles
$(\theta_x, \theta_y, \theta_z)$ and the mass ratio
$\xi \equiv m_2/m_3 \approx 0.16$ \cite{GX03}:
\begin{eqnarray}
\delta & = & \arccos \left [ \frac{c^2_y s^2_z - \xi^2 s^2_x
\left (c^2_x s^2_y + s^2_x c^2_y s^2_z \right )}{2 \xi^2 s^3_x c_x
s_y c_y s_z} \right ] \; ,
\nonumber \\
\sigma & = & \frac{1}{2} \arctan \left [\frac{c_x s_y \sin\delta}
{s_x c_y s_z + c_x s_y \cos\delta} \right ] \; .
%       (16)
\end{eqnarray}
Because $|\cos\delta| \leq 1$ must hold, we find that $\theta_z$
is restricted to a very narrow range:
$4.0^\circ \lesssim \theta_z \lesssim 4.4^\circ$ (i.e.,
$0.070 \lesssim s_z \lesssim 0.077$). The implication of this result
is that the FGY ansatz with $m_1 =0$ will simply be excluded, if the
experimental value of $\theta_z$ does not really lie in the predicted
region.

As a direct consequence of $m_1 =0$, the RGEs of $m_1$, $m_2$ and $m_3$
in Eq. (A1) may be simplified to
\begin{eqnarray}
\dot{m}_1 & = & 0 \; ,
\nonumber \\
\dot{m}_2 & \approx & \frac{1}{16 \pi^2} \left( \alpha + 2 C f_\tau^2
c_x^2 s_y^2 \right) m_2 \; ,
\nonumber \\
\dot{m}_3 & \approx & \frac{1}{16 \pi^2} \left( \alpha + 2 C f_\tau^2
c_y^2 \right) m_3 \; .
%       (17)
\end{eqnarray}
One can see that $m_1 =0$ holds
at any energy scale between $M_Z$ and $M_1$, and the running
behaviors of $m_2$ and $m_3$ are essentially identical (dominated
by the term proportional to $\alpha$). To illustrate,
we show the ratio $R \equiv m_2(M_Z)/m_2(M_1)$ changing with
$m^{~}_H$ in the SM or with $\tan\beta$ in the MSSM in Fig. 2,
where $M_1 = 10^{14}$ GeV is typically taken. It becomes clear that
$R_{m_1 =0} \approx I_\alpha$ is an excellent approximation in the
SM, and it is also a good approximation in the MSSM.

We remark that
$m_2/m_3$ is approximately unchanged in the RGE evolution from $M_Z$
to $M_1$ --- in other words, $\xi \approx 0.16$ is nearly
a constant. It is then possible to simplify the RGEs of
$(\theta_x, \theta_y, \theta_z)$ and $(\delta, \sigma)$ in
Eqs. (A2) and (A3) up to ${\cal O}(\xi)$ or ${\cal O}(s_z)$:
\begin{eqnarray}
\dot{\theta}_x & \approx & - \frac{C f_\tau^2}{16 \pi^2} s_x c_x s_y^2 \; ,
\nonumber \\
\dot{\theta}_y & \approx & - \frac{C f_\tau^2}{16 \pi^2} s_y c_y
\left (1 + 2 \xi c_x^2 \cos \delta \right ) \; ,
\nonumber \\
\dot{\theta}_z & \approx & - \frac{C f_\tau^2}{8 \pi^2} \xi s_x c_x s_y
c_y \; ;
%       (18)
\end{eqnarray}
and
\begin{eqnarray}
\dot{\sigma} & \approx & \frac{C f_\tau^2}{8 \pi^2} \left (s_x c_x s_y
c_y \frac{\xi}{s_z} \right ) \xi \sin \delta \; ,
\nonumber \\
\dot{\delta} & \approx & \frac{C f_\tau^2}{8 \pi^2}
\left [s_x c_x s_y c_y \frac{\xi}{s_z} + c^2_x \left( c_y^2 - s_y^2 \right)
\right] \xi \sin\delta \; .
%       (19)
\end{eqnarray}
In obtaining Eqs. (18) and (19), we have considered the fact that
$\xi \sim 2 s_z$ holds at $M_Z$. One can see that the running effects
of three mixing angles and two CP-violating phases are all governed by
$f^2_\tau$. Because of $f^2_\tau \approx 10^{-4}$ in the SM, the
evolution of $(\theta_x, \theta_y, \theta_z)$ and $(\sigma, \delta)$
is negligibly small. When $\tan\beta$ is sufficiently large (e.g.,
$\tan\beta \sim 50$) in the MSSM, however,
$f^2_\tau \approx 10^{-4} /\cos^2\beta$ can be of ${\cal O}(0.1)$ and
even close to unity --- in this case, some small variation of
$(\theta_x, \theta_y, \theta_z)$ and $(\sigma, \delta)$ due to the RGE
running from $M_Z$ to $M_1$ will appear. Let us define
$\Delta \theta_i \equiv \theta_i (M_1) - \theta_i (M_Z)$
(for $i=x,y,z$),
$\Delta \delta \equiv \delta (M_1) - \delta (M_Z)$ and
$\Delta \sigma \equiv \sigma (M_1) - \sigma (M_Z)$.
The numerical results of $\Delta \theta_i/\theta_i (M_Z)$,
$\Delta \delta/\delta (M_Z)$ and $\Delta \sigma/\sigma (M_Z)$
are shown in Fig. 3 for the MSSM with different values of $M_1$ and
$\tan\beta$. We see that the ratio $\Delta \theta_z/\theta_z (M_Z)$
is most sensitive to the RGE running, but its magnitude is less than
$10\%$ even if $M_1 = 10^{14}$ GeV and $\tan\beta = 50$ are taken.
Thus we conclude that the RGE effects on three flavor mixing angles
and two CP-violating phases are practically negligible in the FGY
ansatz with $m_1 =0$.

\subsection{Inverted neutrino mass hierarchy ($m_3 =0$)}

If $m_3 =0$ holds, we will arrive at
$m_1 =\sqrt{\Delta m^2_{\rm atm} - \Delta m^2_{\rm sun}}
\approx 5.0 \times 10^{-2}$ eV and $m_2 =\sqrt{\Delta m^2_{\rm atm}}
\approx 5.1 \times 10^{-2}$ eV. In this case, only the difference of
two Majorana CP-violating phases $\sigma -\rho \equiv \sigma'$
is physically meaningful. Again, the texture zero $(M_\nu)_{13} =0$
allows us to determine $\delta$ and $\sigma'$ in terms of
the flavor mixing angles $(\theta_x, \theta_y, \theta_z)$ and the mass
ratio $\zeta \equiv m_1/m_2 \approx 0.98$ \cite{GX03}:
\begin{eqnarray}
\delta & = & \arccos \left [ \frac{\left (\zeta^2 c^4_x -
s^4_x \right ) c^2_y s^2_z + \left (\zeta^2 - 1 \right ) s^2_x
c^2_x s^2_y}{2 s_x c_x \left (s^2_x + \zeta^2 c^2_x \right )
s_y c_y s_z} \right ] \; ,
\nonumber \\
\sigma' & = & -\frac{1}{2}
\arctan \left [\frac{s_y c_y s_z \sin\delta}
{s_x c_x \left (s^2_y - c^2_y s^2_z \right ) +
\left (s^2_x - c^2_x \right ) s_y c_y s_z \cos\delta} \right ] \; .
%       (20)
\end{eqnarray}
Different from the $m_1 =0$ case, $|\cos\delta| \leq 1$ does not
impose significant constraints on the magnitude of $\theta_z$
via Eq. (20), except that it requires
$\theta_z > 0.36^\circ$ \cite{GX03}. Hence the FGY ansatz with $m_3 =0$
is not very sensitive to the measurement of $\theta_z$.

As a straightforward consequence of $m_3 =0$, the RGEs of $m_1$,
$m_2$ and $m_3$ in Eq. (A1) can be simplified to
\begin{eqnarray}
\dot{m}_1 & \approx & \frac{1}{16 \pi^2} \left (\alpha + 2 C f_\tau^2
s_x^2 s_y^2 \right ) m_1  \; ,
\nonumber \\
\dot{m}_2 & \approx & \frac{1}{16 \pi^2} \left (\alpha + 2 C f_\tau^2
c_x^2 s_y^2 \right ) m_2  \; ,
\nonumber \\
\dot{m}_3 & = & 0 \; .
%       (21)
\end{eqnarray}
Again, $m_3 =0$ holds at any
energy scale between $M_Z$ and $M_1$; and the running behaviors of
$m_1$ and $m_2$ are essentially the same. Fig. 2 illustrates the
ratio $R \equiv m_2(M_Z)/m_2(M_1)$ as a function of $m^{~}_H$
in the SM or of $\tan\beta$ in the MSSM. We see that
$R_{m_3 =0} \approx R_{m_1 =0} \approx I_\alpha$ holds to an
excellent degree of accuracy in the SM and to a good degree of
accuracy in the MSSM. These numerical results confirm that the
evolution of three neutrino masses is dominated by $I_\alpha$,
as observed in section II.

While $\zeta$ is approximately a constant in the RGE running
from $M_Z$ to $M_1$, it is not small. In this case, we simplify
Eqs. (A2) and (A3) up to ${\cal O}(s_z)$ so as to get the
leading-order RGEs of $(\theta_x, \theta_y, \theta_z)$ and
$(\delta, \sigma')$ as follows:
\begin{eqnarray}
\dot{\theta}_x & \approx & - \frac{C f_\tau^2}{16 \pi^2}
\left( \frac{ 1 + \zeta^2 + 2 \zeta \cos 2 \sigma'}
{1 - \zeta^2} \right) s_x c_x s_y^2 \; ,
\nonumber \\
\dot{\theta}_y & \approx & \frac{C f_\tau^2}{16 \pi^2}
s_y c_y \; ,
\nonumber \\
\dot{\theta}_z & \approx & \frac{C f_\tau^2}{16 \pi^2} c_y^2 s_z \; ;
%       (22)
\end{eqnarray}
and
\begin{eqnarray}
\dot{\sigma}' & \approx & \frac{C f_\tau^2}{8 \pi^2}
\left ( \frac{c_x^2 - s_x^2}{1 - \zeta^2} \right )
\zeta s^2_y \sin 2 \sigma' \; ,
\nonumber \\
\dot{\delta} & \approx & \frac{C f_\tau^2}{8 \pi^2}
\left (\frac{1}{1 - \zeta^2} \right )
\zeta s_y^2 \sin 2 \sigma' \; .
%       (23)
\end{eqnarray}
Unlike the $m_1 =0$ case, the RGE running effects of $\theta_x$,
$\delta$ and $\sigma'$ are enhanced by a factor
$1/(1-\zeta^2) \approx 25$ in the $m_3 =0$ case. One might
naively expect that the magnitudes of
$\Delta \theta_x \equiv \theta_x (M_1) - \theta_x (M_Z)$,
$\Delta \delta \equiv \delta (M_1) - \delta (M_Z)$ and
$\Delta \sigma' \equiv \sigma' (M_1) -
\sigma' (M_Z)$ are appreciable. Because of
$\sin \sigma' \sim {\cal O}(s_z)$,
however, $\dot{\delta}$ and $\dot{\sigma}'$ in
Eq. (23) are actually suppressed. Therefore, only $\theta_x$
is likely to be sensitive to the RGE evolution from $M_Z$
to $M_1$. We plot the numerical results of
$\Delta \theta_i/\theta_i (M_Z)$ (for $i=x,y,z$),
$\Delta \delta/\delta (M_Z)$ and
$\Delta \sigma'/\sigma' (M_Z)$
in Fig. 4 for the SM and in Fig. 5 for the MSSM.
One can see that the RGE effects on three flavor mixing angles
and two CP-violating phases are negligibly small in the SM,
but they may become significant in the MSSM if both $M_1$
and $\tan\beta$ are sufficiently large. In the latter case,
$\theta_x (M_1)$ is even possible to approach zero --- this
point has indeed been observed by some authors beyond the
minimal seesaw model \cite{ThetaX}. We conclude that the
near degeneracy between $m_1$ and $m_2$ in the $m_3 =0$
case may give rise to significant RGE running effects on
the mixing angle $\theta_x$ in the MSSM, and the evolution
of CP-violating phases $\delta$ and $\sigma'$
can also be appreciable if both $M_1$ and $\tan\beta$ take
properly large values.

\section{Cosmological baryon number asymmetry}

Lepton number violation induced by the third term of
${\cal L}_{\rm Y}$ in Eq. (1) or Eq. (2) allows decays of
the heavy Majorana neutrinos $N_i$
(for $i=1$ and 2) to happen: $N_i \rightarrow l + h$ and
$N_i \rightarrow \bar{l} + h^{\rm c}$, where $h = H$ in the SM or
$h = H^{\rm c}_2$ in the MSSM. Because each decay mode occurs at
both tree and one-loop levels (via the self-energy and vertex
corrections), the interference between these two decay amplitudes may
result in a CP-violating asymmetry $\varepsilon_i$ between
$N_i \rightarrow l + h$ and its $CP$-conjugated process \cite{FY}.
If the masses of $N_1$ and $N_2$ are hierarchical
(i.e., $M_1 \ll M_2$), the interactions of $N_1$ can be in thermal
equilibrium when $N_2$ decays. The asymmetry $\varepsilon_2$
is therefore erased before $N_1$ decays, and only the asymmetry
$\varepsilon_1$ produced by the out-of-equilibrium decay
of $N_1$ survives. In the flavor basis chosen above, we have
\begin{eqnarray}
\varepsilon_1 & \equiv & \frac{\Gamma (N_1 \rightarrow l + h)
~ - ~ \Gamma (N_1 \rightarrow \bar{l} + h^{\rm c})}
 {\Gamma (N_1 \rightarrow l + h)
~ + ~ \Gamma (N_1 \rightarrow \bar{l} + h^{\rm c})}
\nonumber \\
& \approx & \frac{C'}{8\pi} \cdot \frac{M_1}{M_2} \cdot
\frac{{\rm Im} \left [ (Y^\dagger_\nu Y_\nu)_{12} \right ]^2}
{(Y^\dagger_\nu Y_\nu)_{11}} \;\; ,
%       (24)
\end{eqnarray}
where $C'_{\rm SM} = -3/2$ and $C'_{\rm MSSM} = -3$ \cite{R}.
Taking account of Eq. (3) with $a_2 = c_1 =0$, we immediately
arrive at
\begin{eqnarray}
(Y^\dagger_\nu Y_\nu)_{11} & = & |a_1(M_1)|^2 + |b_1(M_1)|^2
\nonumber \\
& \approx  & \frac{1}{I_\alpha} \left [ |a_1(M_Z)|^2 + |b_1 (M_Z)|^2
\right ] \; ,
\nonumber \\
(Y^\dagger_\nu Y_\nu)_{12} & = & b^*_1(M_1) \cdot b_2(M_1)
\nonumber \\
& \approx & \frac{1}{I_\alpha} \left [ b^*_1(M_Z) \cdot b_2(M_Z)
\right ] \; ,
%   (25)
\end{eqnarray}
where $I_e \approx I_\mu \approx 1$ has been used as an excellent
approximation.

In the literature, $\varepsilon_1$ was calculated by neglecting
the RGE running effects of neutrino masses, lepton flavor mixing
angles and CP-violating phases from $M_Z$ to $M_1$ (i.e.,
$I_\alpha \approx 1$ was naively taken). Such an oversimplification
leads to the CP-violating asymmetry
\begin{equation}
\hat{\varepsilon}_1 \; \approx \;
\frac{C'}{8\pi \Omega} \cdot
\frac{\displaystyle M_1 |(M_\nu)_{12}|^2 |(M_\nu)_{23}|^2}
{\displaystyle \left [ |(M_\nu)_{11}|^2 + |(M_\nu)_{12}|^2 \right ]
|(M_\nu)_{33}|} \sin\Phi \; ,
%       (26)
\end{equation}
where $\Omega = v^2$ (SM) or $\Omega = v^2 \sin^2\beta$ (MSSM),
and \cite{GX03}
\begin{equation}
\Phi \; \approx \; \left \{ \matrix{ \displaystyle \arctan \left (
\frac{\xi s^2_x c^2_z \sin 2\sigma}{s^2_z + \xi s^2_x c^2_z \cos
2\sigma} \right ) \; , ~~ (m_1 =0) \; , ~~~ \cr\cr \displaystyle -
2 \delta  \; , ~~ (m_3 = 0) \; . ~~~~~~~~~~~~~~~~~~~~~~~~~~~~~~~~~
\cr} \right .
%       (27)
\end{equation}
One can see that $\hat{\varepsilon}_1$ is actually independent of $M_2$,
as long as $M_2 \gg M_1$ is satisfied. Then it is straightforward to
obtain $\varepsilon_1 \approx \hat{\varepsilon}_1/I_\alpha$ at the
mass scale $M_1$. Although $I_\alpha$ is always smaller than unity for
$M_1 > M_Z$ (as already shown in Fig. 1), it remains of ${\cal O}(1)$
only if reasonable values of $M_1$ and $m^{~}_H$ (SM) or
$\tan\beta$ (MSSM) are taken. Hence the previously
oversimplified approximation $\varepsilon_1 \approx \hat{\varepsilon}_1$
is unable to cause any quantitative disaster.

A nonvanishing CP-violating asymmetry $\varepsilon_1$ may result in
a net lepton number asymmetry $Y_{\rm L} \equiv n^{~}_{\rm L}/{\bf s} =
\varepsilon_1 d/g^{~}_*$, where $g^{~}_* = 106.75$ (SM) or 228.75 (MSSM)
is an effective number characterizing the relativistic degrees of freedom
which contribute to the entropy {\bf s} of the early universe, and $d$
accounts for the dilution effects induced by the lepton-number-violating
wash-out processes \cite{R}. If the effective neutrino mass parameter
$\tilde{m}_1 \equiv (Y^\dagger_\nu Y_\nu)_{11} \Omega/M_1$ \cite{BP}
lies in the range
$10^{-2} ~ {\rm eV} \leq \tilde{m}_1 \leq 10^3 ~ {\rm eV}$, one may
estimate the value of $d$ by using the following approximate
formula \cite{Kolb}
%%%%%%%%%%%%%%%%%%
\footnote{In view of Eq. (25), we can easily obtain
$\tilde{m}_1 (M_1) \approx \tilde{m}_1 (M_Z)/I_\alpha$. It is
proper to use $\tilde{m}_1 (M_1)$ instead of
$\tilde{m}_1 (M_Z)$ to evaluate the dilution factor $d$ via
Eq. (28), but the numerical discrepancy between
$\tilde{m}_1 (M_1)$ and $\tilde{m}_1 (M_Z)$ is actually insignificant.}:
%%%%%%%%%%%%%%%%%
\begin{equation}
d \; \approx \; 0.3 \left (\frac{10^{-3} ~ {\rm eV}}{\tilde{m}_1} \right )
\left [ \ln \left ( \frac{\tilde{m}_1}{10^{-3} ~ {\rm eV}} \right )
\right ]^{-0.6} \; . ~~~
%       (28)
\end{equation}
The lepton number asymmetry
$Y_{\rm L}$ is eventually converted into a net baryon number asymmetry
$Y_{\rm B}$ via the nonperturbative sphaleron processes \cite{Kuzmin}:
$Y_{\rm B} \equiv n^{~}_{\rm B}/{\bf s} \approx -0.55 Y_{\rm L}$ in the
SM or $Y_{\rm B} \equiv n^{~}_{\rm B}/{\bf s} \approx -0.53 Y_{\rm L}$
in the MSSM. A generous range
$0.7 \times 10^{-10} \lesssim Y_{\rm B} \lesssim 1.0 \times 10^{-10}$
has been drawn from the recent WMAP observational data \cite{WMAP}.

Clearly $\varepsilon_1$ and $Y_{\rm B}$ involve three free
parameters in the SM ($M_1$, $\theta_z$ and $m^{~}_H$) and
three free parameters in the MSSM ($M_1$, $\theta_z$ and $\tan\beta$).
Note, however, that $I_\alpha$ is not very sensitive to $m^{~}_H$
in the SM and its sensitivity to $\tan\beta$ is mild in the MSSM
(see Fig. 1 for illustration). Therefore the size of $Y_{\rm B}$
calculated at $M_1$ is mainly dependent on how big $M_1$ and $\theta_z$
are. The numerical dependence of $Y_{\rm B}$ on $M_1$ and $\theta_z$
is illustrated in Fig. 6, where $m^{~}_H = 120$ GeV (SM) and
$\tan\beta = 50$ (MSSM) have typically been taken.
Some comments are in order.

(1) In the $m_1 =0$ case, current observational data of $Y_{\rm B}$
require $M_1 > 3.1 \times 10^{10}$ GeV (SM) or
$M_1 > 3.4 \times 10^{10}$ GeV (MSSM) for the allowed range of
$\theta_z (M_Z)$. Once this mixing angle is precisely measured at
low energies, it is possible to fix the ball-park magnitude of
$M_1$ in most cases. However, to distinguish between the minimal
seesaw SM and its supersymmetric version needs other experimental
information (e.g., the MSSM-motivated lepton-flavor-violating
processes $\mu \rightarrow e\gamma$, $\tau \rightarrow \mu\gamma$,
and so on \cite{Raidal}).

(2) In the $m_3 =0$ case, $M_1 > 2.25 \times 10^{13}$ GeV (SM)
or $M_1 > 2.5 \times 10^{13}$ GeV (MSSM)
is required by current observational
data of $Y_{\rm B}$. The magnitude of $Y_{\rm B}$ increases
monotonically with $\theta_z (M_Z)$ for any given value of $M_1$.
Once $\theta_z$ is measured at low energies, it is possible to
determine the rough value of $M_1$. Again, other experimental
information is needed in order to distinguish between the minimal
seesaw SM and its supersymmetric version.

(3) We remark that there is a potential conflict between achieving
successful thermal leptogenesis and avoiding overproduction of
gravitinos in the minimal seesaw model with supersymmetry \cite{Raidal}.
If the mass scale of gravitinos is of ${\cal O}(1)$ TeV, one must
have $M_1 \lesssim 10^8$ GeV. This limit is completely disfavored in
the FGY ansatz with $M_1 \ll M_2$, because it would lead to
$Y_{\rm B} \ll 10^{-10}$. If $M_1$ and $M_2$ were almost degenerate,
a special case which has been discussed in Ref. \cite{R},
it would be possible to simultaneously accommodate
$m^{~}_{\tilde G} \sim {\cal O}(1)$ TeV and $M_1 \lesssim 10^8$ GeV
in the generic supergravity models with minimal seesaw and thermal
leptogenesis.

\section{Summary}

We have analyzed the radiative corrections to neutrino mixing and CP
violation in the minimal seesaw model with two heavy right-handed
neutrinos. It is shown that textures and vanishing eigenvalues
of the effective Majorana neutrino mass matrix
are essentially stable against renormalization effects. Taking account
of the Frampton-Glashow-Yanagida ansatz for the Dirac neutrino Yukawa
coupling matrix, we have calculated the RGE running effects of light
neutrino masses, lepton flavor mixing angles and CP-violating phases
from $M_Z$ to $M_1$ for both $m_1 =0$ and $m_3 =0$ cases in the SM
and its minimal supersymmetric extension. We find that such
quantum corrections are not always negligible, and they should
be taken into consideration in order to quantitatively test the FGY
ansatz. We have also discussed thermal leptogenesis in the minimal seesaw
model with $M_1 \ll M_2$. Very instructive predictions for the cosmological
baryon number asymmetry are obtained with the help of low-energy neutrino
mixing quantities. We conclude that a precise measurement of the mixing
angle $\theta_z$ in reactor- and accelerator-based neutrino oscillation
experiments will be extremely helpful to examine the FGY scenario and other
presently viable ans$\rm\ddot{a}$tze of lepton mass matrices.

\acknowledgments{
We are grateful to W.L. Guo for very useful
discussions and helps. This work was supported in
part by the National Nature Science Foundation of China.}

\newpage

\appendix
\section{Analytical approximations of the RGEs for neutrino masses
and lepton flavor mixing parameters}
\setcounter{equation}{0}

Following Ref. \cite{Casas} and taking account of $M_\nu = V {\rm
Diag}\{m_1, m_2, m_3\} V^T$ with the parametrization of $V$ given
in Eq. (12), we have derived the one-loop RGEs of $(m_1, m_2,
m_3)$, $(\theta_x, \theta_y, \theta_z)$ and $(\delta, \rho,
\sigma)$ with the help of Eq. (5). Our analytical results are
consistent with those obtained in Ref. \cite{Lindner}, where a
somehow different parametrization of $V$ has been used. For
simplicity, only the approximate expressions of $(\dot{m}_1,
\dot{m}_2, \dot{m}_3)$, $(\dot{\theta}_x, \dot{\theta}_y,
\dot{\theta}_z)$ and $(\dot{\delta}, \dot{\rho}, \dot{\sigma})$ up
to ${\cal O}(s_z)$ are presented below.

(1) The running of three neutrino masses:
\begin{eqnarray}
\dot{m}_1 &=& \frac{1}{16 \pi^2} \left[ \alpha + 2 C f_\tau^2
\left( s_x^2 s_y^2 + {\cal O} (s_z) \right) \right] m_1  \; ,
\nonumber \\
\dot{m}_2 &=& \frac{1}{16 \pi^2} \left[ \alpha + 2 C f_\tau^2
\left( c_x^2 s_y^2 + {\cal O} (s_z) \right) \right] m_2 \; ,
\nonumber \\
\dot{m}_3 &=& \frac{1}{16 \pi^2} \left[ \alpha + 2 C f_\tau^2
c_y^2 c_z^2 \right] m_3 \; .
%       (A.1)
\end{eqnarray}

(2) The running of three lepton flavor mixing angles:
\begin{eqnarray}
\dot{\theta}_x &=& - \frac{C f_\tau^2}{16 \pi^2} \left[ \frac{
m_1^2 + m_2^2 +2 m_1 m_2 \cos 2 (\rho - \sigma) }{m_2^2 -
m_1^2} s_x c_x s_y^2  + {\cal O} (s_z) \right] \; ,
\nonumber \\
\dot{\theta}_y &=& - \frac{C f_\tau^2}{16 \pi^2} \left[
\frac{m_2^2 + m_3^2 + 2 m_2 m_3 \cos 2(\delta - \sigma)}{m_3^2 -
m_2^2} c^2_x s_y c_y \right .
\nonumber \\
& & \; \; \; \; \; \; \; \; \; \; \; \;
\left . + \frac{m_1^2 + m_3^2 + 2 m_1 m_3 \cos 2(\delta
- \rho) }{m_3^2 - m_1^2} s^2_x s_y c_y  + {\cal O} (s_z) \right] \; ,
\nonumber \\
\dot{\theta}_z &=& - \frac{C f_\tau^2}{16 \pi^2} \left[ \left (
\frac{2 m_2 m_3 \cos(\delta - 2 \sigma)} {m_3^2 - m_2^2} - \frac{2
m_1 m_3 \cos(\delta - 2 \rho)} {m_3^2 - m_1^2} \right . \right .
\nonumber \\
& & \; \; \; \; \; \; \; \; \; \; \;
\left . \left . + \frac{m_2^2 - m_1^2}{m_3^2 - m_2^2} \cdot
\frac{2 m_3^2 \cos \delta}{m_3^2 - m_1^2} \right ) s_x c_x s_y c_y
+ {\cal O} (s_z) \right] \; .
%       (A.2)
\end{eqnarray}

(3) The running of three CP-violating phases:
\begin{eqnarray}
\dot{\rho} &=& - \frac{C f_\tau^2}{16 \pi^2} \left[ \frac{ A}{
s_z} - \frac{2 m_2 m_3 \sin 2 \sigma}{m_3^2 - m_2^2} s_x^2 c_y^2 -
\frac{2 m_1 m_3 \sin 2 \rho}{m_3^2 - m_1^2} c_x^2 c_y^2 \right .
\nonumber \\
& & \; \; \; \; \; \; \; \; \; \; \; \;
\left . + \frac{2 m_1 m_2 \sin 2 (\rho - \sigma)}{m_2^2 - m_1^2} s_x^2
s_y^2 + {\cal O} (s_z) \right] \; ,
\nonumber \\
\dot{\sigma} &=& - \frac{C f_\tau^2}{16 \pi^2} \left[ \frac{ A}{
s_z} - \frac{2 m_2 m_3 \sin 2 \sigma}{m_3^2 - m_2^2} s_x^2 c_y^2 -
\frac{2 m_1 m_3 \sin 2 \rho}{m_3^2 - m_1^2} c_x^2 c_y^2 \right .
\nonumber \\
& & \; \; \; \; \; \; \; \; \; \; \; \;
\left . + \frac{2 m_1 m_2 \sin 2(\rho - \sigma)}{m_2^2 -
m_1^2} c^2_x s_y^2 + {\cal O} (s_z) \right] \; ,
\nonumber \\
\dot{\delta} &=& - \frac{C f_\tau^2}{16 \pi^2} \left[ \frac{A}
{s_z} + B + {\cal O} (s_z) \right] \; ,
%       (A.3)
\end{eqnarray}
where
\begin{eqnarray}
A &=& s_x c_x s_y c_y \left[ \frac{2 m_2 m_3 \sin (\delta - 2
\sigma)}{m_3^2 - m_2^2} - \frac{2 m_1 m_3 \sin (\delta - 2
\rho)}{m_3^2 - m_1^2} \right .
\nonumber \\
& & \; \; \; \; \; \;
\; \; \; \; \; \; \; \; \left . - \frac{m_2^2 - m_1^2}{m_3^2 - m_2^2}
\cdot \frac{2 m_3^2 \sin \delta }{m_3^2 - m_1^2} \right] \; ,
\nonumber \\
B &=& \frac{2 m_1 m_2 s^2_y \sin 2(\rho - \sigma)}{m_2^2 - m_1^2}
\nonumber \\
& & - \frac{2 m_2 m_3}{m_3^2 - m_2^2} \left [
s_x^2 c^2_y \sin 2\sigma + c_x^2 \left (c_y^2 -s_y^2 \right )
\sin 2(\delta - \sigma) \right ]
\nonumber \\
& & - \frac{2 m_1 m_3}{m_3^2 - m_1^2}
\left [ c_x^2 c_y^2 \sin 2\rho + s^2_x \left (c_y^2 -s_y^2 \right )
\sin 2(\delta - \rho) \right ] \; .
\end{eqnarray}

\newpage

%%%%%%%%%%%%%%%%%%%% Fig. 1 %%%%%%%%%%%%%%%%
\begin{figure}[t]
\vspace{2cm}
\epsfig{file=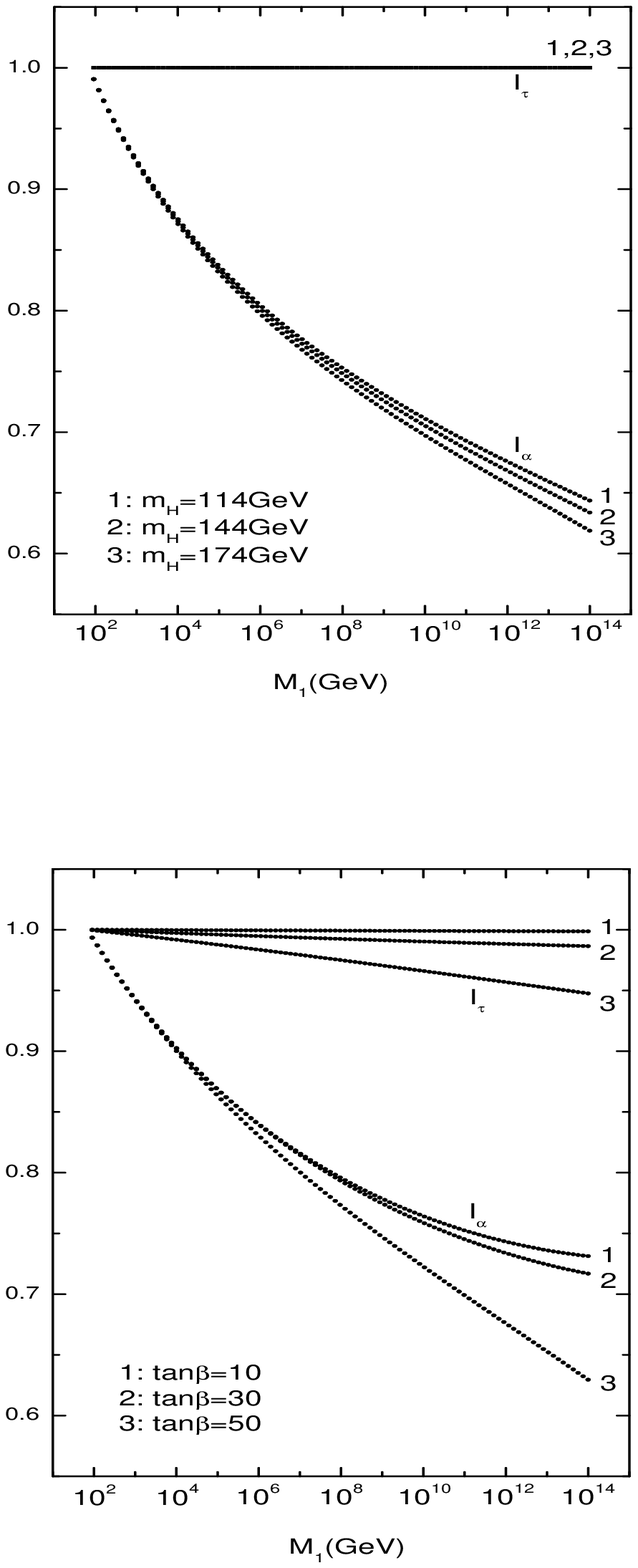,bbllx=1cm,bblly=4cm,bburx=20cm,bbury=26cm,
width=14.5cm,height=17cm,angle=0,clip=0}
\vspace{0.4cm}
\caption{Numerical illustration of the evolution functions
$ I_\alpha$ and $I_\tau $ changing with $M_1$ and $m^{~}_H$
in the SM (up) or with $M_1$ and $\tan\beta$ in the MSSM (down).}
\end{figure}

%%%%%%%%%%%%%%%%%%%% Fig. 2 %%%%%%%%%%%%%%%%
\begin{figure}[t]
\vspace{2cm}
\epsfig{file=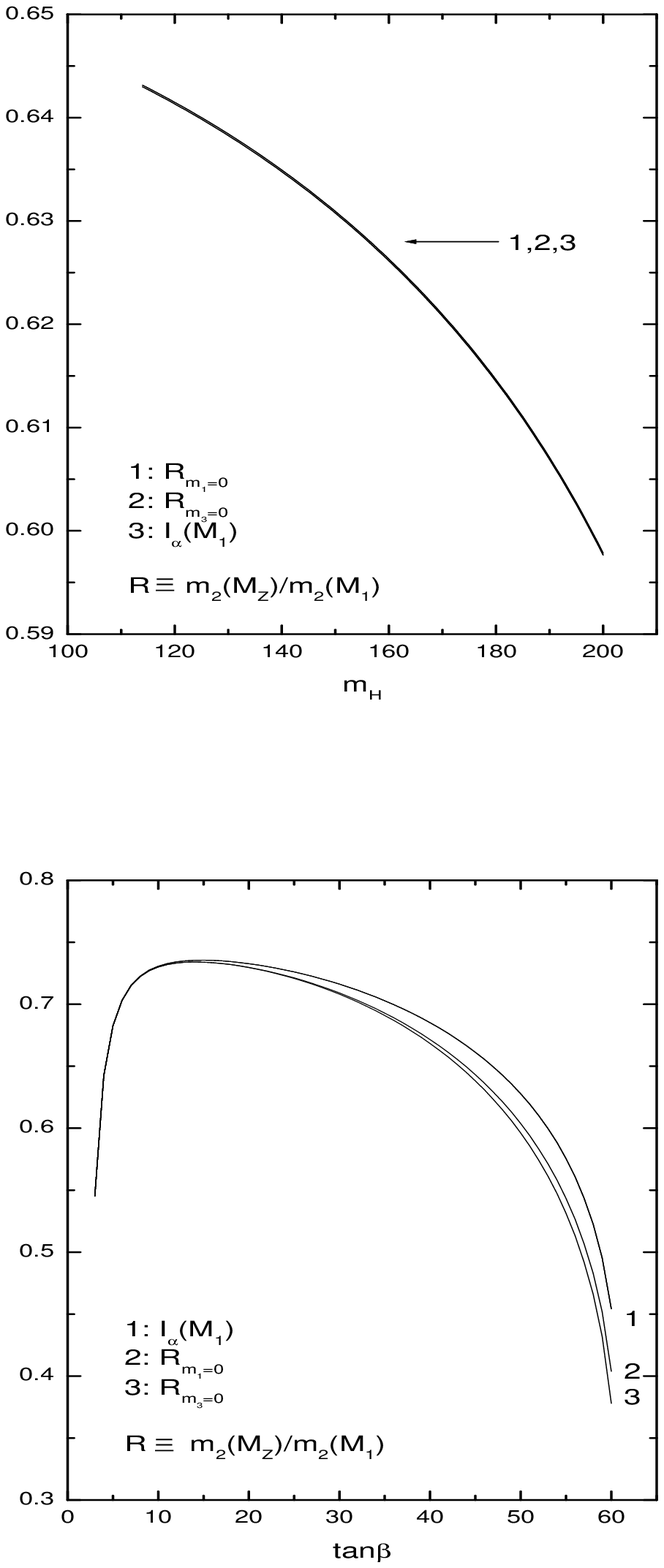,bbllx=1cm,bblly=4cm,bburx=20cm,bbury=26cm,
width=14.5cm,height=17cm,angle=0,clip=0}
\vspace{0.3cm}
\caption{The ratio of $m_2$ at $M_Z$ to its value at $M_1= 10^{14}$ GeV
as a function of $m^{~}_H$ in the SM (up) or of $\tan\beta$ in the
MSSM (down) for the FGY ansatz with $m_1 =0$ or $m_3 =0$.}
\end{figure}

%%%%%%%%%%%%%%%%%%%% Fig. 3 %%%%%%%%%%%%%%%%
\begin{figure}[t]
\epsfig{file=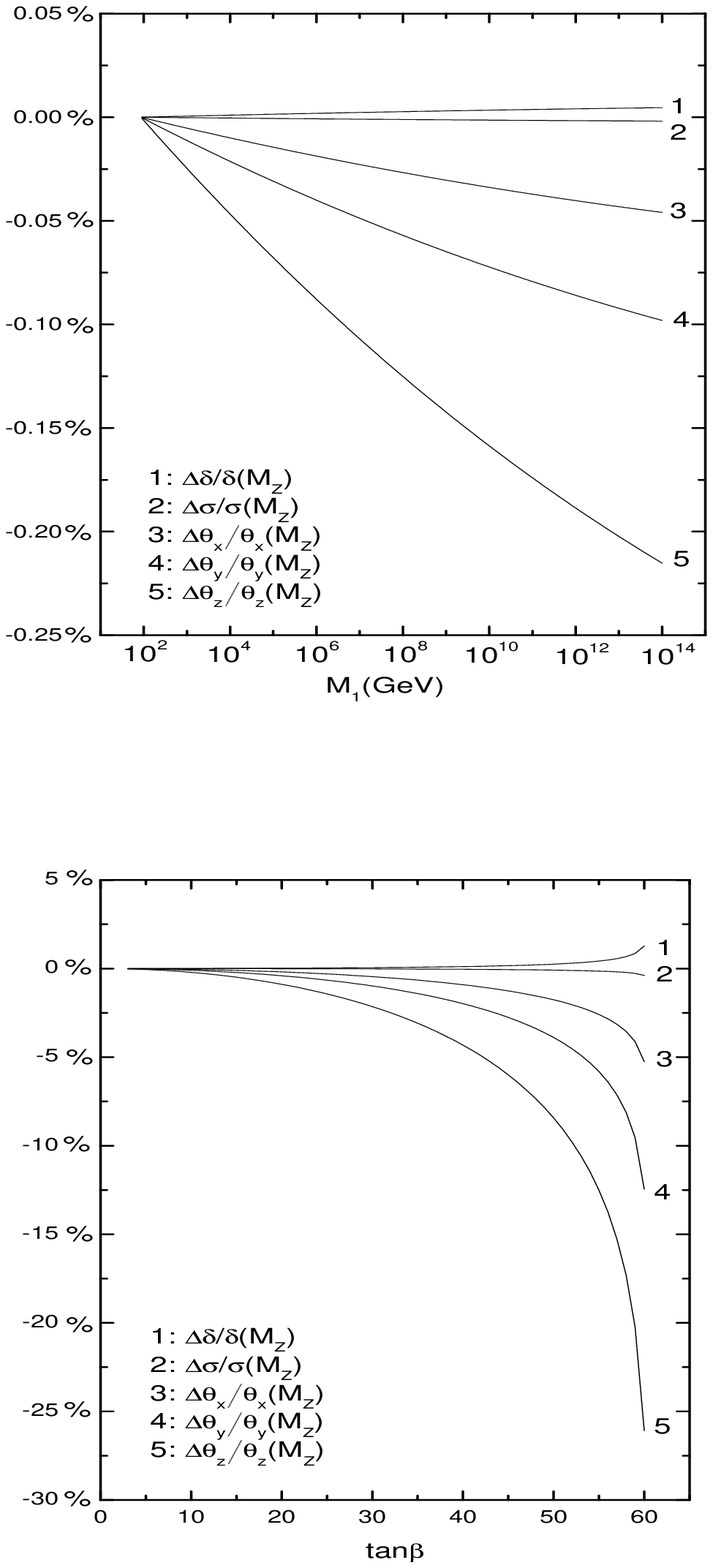,bbllx=1cm,bblly=4cm,bburx=20cm,bbury=26cm,
width=14.5cm,height=17cm,angle=0,clip=0}
\vspace{0.3cm}
\caption{The RGE evolution of lepton flavor mixing angles and
CP-violating phases for the FGY ansatz with $m_1 =0$ in the MSSM.
We take $\tan \beta =10$ to illustrate the running effects changing
with $M_1$ (up), and take $M_1 = 10^{14}$ GeV to illustrate the
running effects changing with $\tan\beta$ (down). Note that
$\theta_z(M_Z) \approx 4.3^\circ$ has typically been input.}
\end{figure}

%%%%%%%%%%%%%%%%%%%% Fig. 4 %%%%%%%%%%%%%%%%
\begin{figure}[t]
\epsfig{file=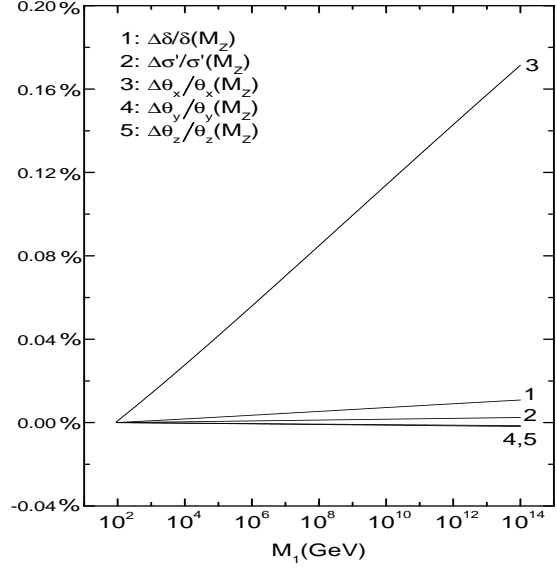,bbllx=1cm,bblly=4cm,bburx=20cm,bbury=26cm,
width=14.5cm,height=17cm,angle=0,clip=0}
\vspace{-4.1cm}
\caption{The RGE evolution of lepton flavor mixing angles and
CP-violating phases from $M_Z$ to $M_1$ for the FGY ansatz with
$m_3 =0$ in the SM. We have typically input
$\theta_z(M_Z) \approx 4.3^\circ$, and found that
the influence of $m^{~}_H$ is negligible.}
\end{figure}

%%%%%%%%%%%%%%%%%%%% Fig. 5 %%%%%%%%%%%%%%%%
\begin{figure}[t]
\epsfig{file=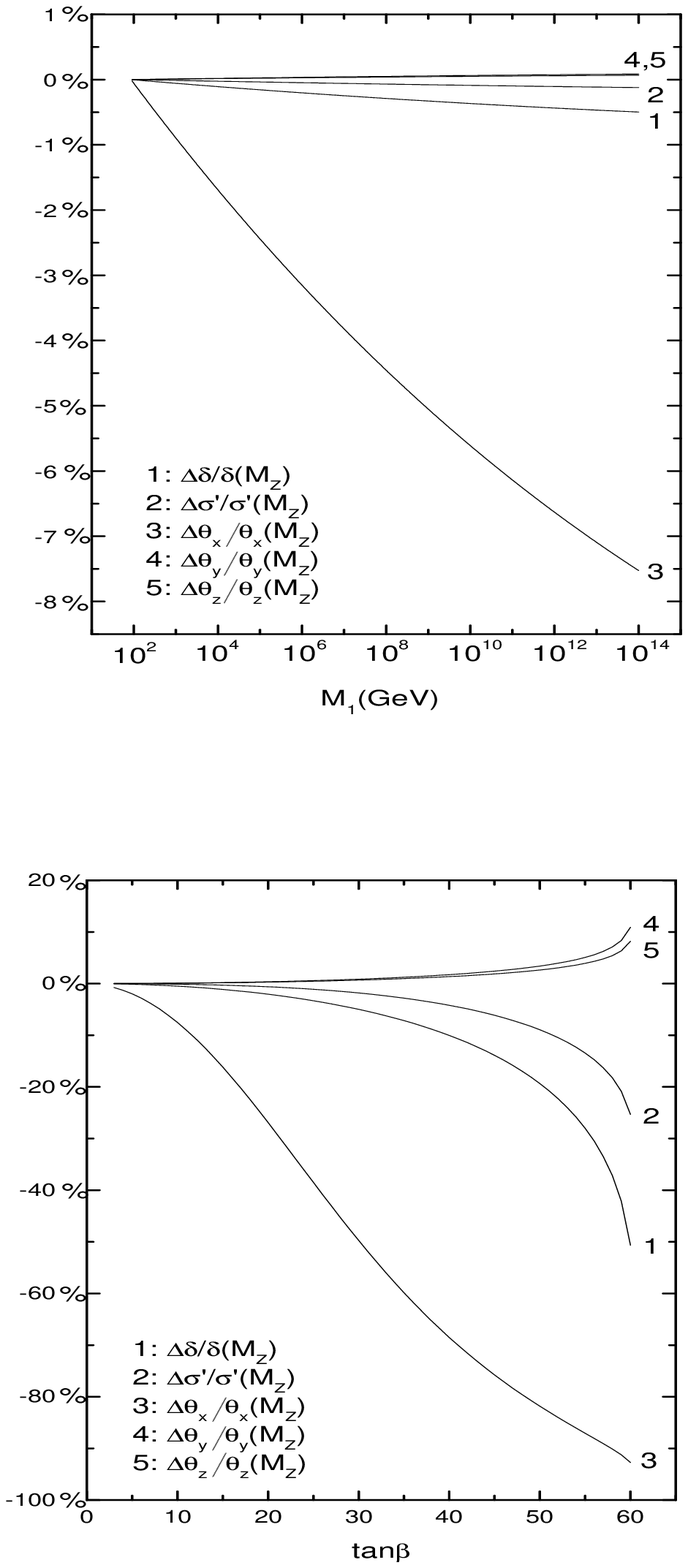,bbllx=1cm,bblly=4cm,bburx=20cm,bbury=26cm,
width=14.5cm,height=17cm,angle=0,clip=0}
\vspace{0.3cm}
\caption{The RGE evolution of lepton flavor mixing angles and
CP-violating phases for the FGY ansatz with $m_3 =0$ in the MSSM.
We take $\tan \beta =10$ to illustrate the running effects changing
with $M_1$ (up), and take $M_1 = 10^{14}$ GeV to illustrate the
running effects changing with $\tan\beta$ (down).
Note that $\theta_z(M_Z) \approx 4.3^\circ$ has typically been input.}
\end{figure}

%%%%%%%%%%%%%%%%%%%% Fig. 6 %%%%%%%%%%%%%%%%
\begin{figure}[t]
\epsfig{file=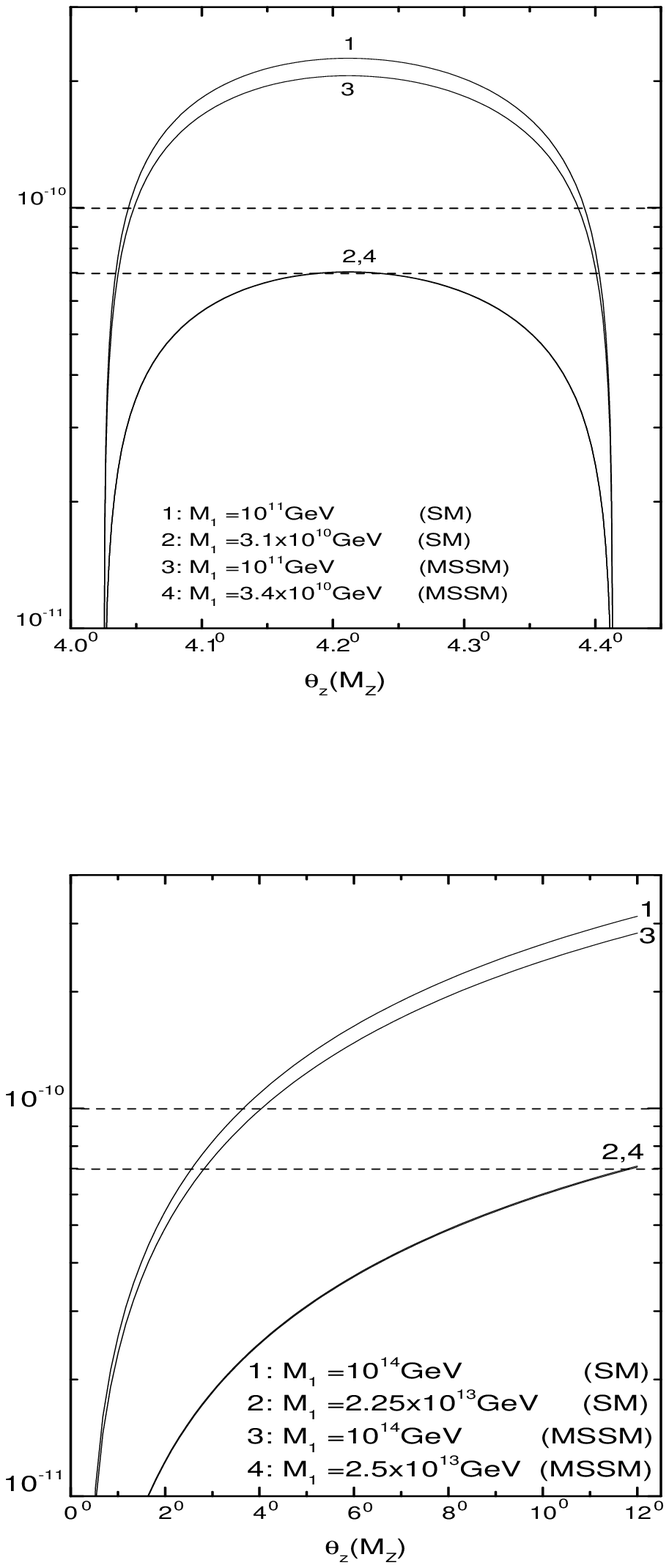,bbllx=1cm,bblly=4cm,bburx=20cm,bbury=26cm,
width=14.5cm,height=17cm,angle=0,clip=0}
\vspace{0.3cm}
\caption{Numerical illustration of $Y_{\rm B}$ changing with
$\theta_z (M_Z)$ for the FGY ansatz with $m_1 =0$ (up) or
$m_3 =0$ (down) in the SM ($m^{~}_H = 120$ GeV)
and its minimal supersymmetric extension ($\tan\beta = 50$).
The region between two dashed lines in each graph corresponding
to the range of $Y_{\rm B}$ allowed by current observational
data.}
\end{figure}

\end{document}